\documentstyle[12pt,graphicx]{article}
\begin{document}
\baselineskip 18pt
\def\today{\ifcase\month\or
 January\or February\or March\or April\or May\or June\or
 July\or August\or September\or October\or November\or December\fi
 \space\number\day, \number\year}

%
\def\thebibliography#1{\section*{References\markboth
 {References}{References}}\list
 {[\arabic{enumi}]}{\settowidth\labelwidth{[#1]}
 \leftmargin\labelwidth
 \advance\leftmargin\labelsep
 \usecounter{enumi}}
 \def\newblock{\hskip .11em plus .33em minus .07em}
 \sloppy
 \sfcode`\.=1000\relax}
\let\endthebibliography=\endlist
\def\beq{\begin{equation}}
\def\eeq{\end{equation}}
\def\beqn{\begin{eqnarray}}
\def\eeqn{\end{eqnarray}}
\def\rmuu{\gamma^{\mu}}
\def\rmud{\gamma_{\mu}}
\def\PL{{1-\gamma_5\over 2}}
\def\PR{{1+\gamma_5\over 2}}
\def\sinW2{\sin^2\theta_W}
\def\AEM{\alpha_{EM}}
\def\mul{M_{\tilde{u} L}^2}
\def\mur{M_{\tilde{u} R}^2}
\def\mdl{M_{\tilde{d} L}^2}
\def\mdr{M_{\tilde{d} R}^2}
\def\mz2{M_{z}^2}
\def\c2b{\cos 2\beta}
\def\au{A_u}
\def\ad{A_d}
\def\cob{\cot \beta}
\def\v#1{v_#1}
\def\tb{\tan\beta}
\def\epem{$e^+e^-$}
\def\KK{$K^0$-$\overline{K^0}$}
\def\wi{\omega_i}
\def\xj{\chi_j}
\def\Wmu{W_\mu}
\def\Wnu{W_\nu}
\def\m#1{{\tilde m}_#1}
\def\mH{m_H}
\def\mw#1{{\tilde m}_{\omega #1}}
\def\mx#1{{\tilde m}_{\chi^{0}_#1}}
\def\mc#1{{\tilde m}_{\chi^{+}_#1}}
\def\mwi{{\tilde m}_{\omega i}}
\def\mxi{{\tilde m}_{\chi^{0}_i}}
\def\mci{{\tilde m}_{\chi^{+}_i}}
\def\mz{M_z}
\def\sw{\sin\theta_W}
\def\cw{\cos\theta_W}
\def\cb{\cos\beta}
\def\sb{\sin\beta}
\def\rwi{r_{\omega i}}
\def\rxj{r_{\chi j}}
\def\rfp{r_f'}
\def\Kik{K_{ik}}
\def\Fq2{F_{2}(q^2)}
\def\f{\({\cal F}\)}
\def\d1{{\f(\tilde c;\tilde s;\tilde W)+ \f(\tilde c;\tilde \mu;\tilde W)}}
\def\tw{\tan\theta_W}
\def\sec2w{sec^2\theta_W}

\begin{titlepage}

\begin{center}
{\Large {\bf  {A Unified Framework for \\[0.1in]
Symmetry Breaking in SO(10)}}}
\end{center}
\begin{center}
\vskip 0.5 true cm \vspace{2cm}
\renewcommand{\thefootnote}
{\fnsymbol{footnote}} {\bf K.S. Babu}$^a$,  {\bf Ilia
Gogoladze}$^b$\footnote{ On leave of absence from Andronikashvili
Institute of Physics, GAS, 380077 Tbilisi, Georgia.}, {\bf Pran Nath}$^{c,d}$ and {\bf Raza M. Syed}$^c$
\vskip 0.5 true cm
\end{center}

\noindent
{a. Department of Physics, Oklahoma State University, Stillwater,
OK, 74078, USA}\\
{b. Department of Physics, University of Notre Dame, Notre Dame,
IN 46556, USA}\\
{c. Department of Physics, Northeastern University,
Boston, MA 02115-5000, USA} \\
{d. Max Planck Institute for Physics, Fohringer Ring 6, D-80805, Munich,
Germany.}\\
\vskip 1.0 true cm
\centerline{\bf Abstract}
\medskip
\noindent

A new SO(10) unified model is proposed based on a one step
breaking of SO(10) to the Standard Model gauge group
$SU(3)_C\times SU(2)_L\times U(1)_Y$ using a single $144$ of
Higgs.
 The symmetry breaking occurs
 when the SU(5) 24-plet component of
$144$ develops a vacuum expectation value. Further, it is possible
to obtain from the same $144$ a light Higgs doublet necessary for
electro-weak symmetry breaking using recent ideas of string vacua
landscapes and fine tuning. Thus the breaking of SO(10) down to
$SU(3)_C\times U(1)_{em}$ can be accomplished with a single Higgs.
We analyze this symmetry breaking pattern in the nonsupersymmetric
as well as in the supersymmetric SO(10) model. In this scenario
masses of the quarks and leptons arise via quartic couplings.  We
show that the resulting mass pattern is consistent with
experimental data, including neutrino oscillations.  The model
represents an alternative to the currently popular grand unified
scenarios.
\end{titlepage}

 \section{Introduction}
 In any Grand Unified Theory (GUT) understanding the Higgs sector is not an easy task. Usually these models
 require the existence of more than one Higgs multiplet. In the minimal $SU(5)$ GUT one employs one adjoint
 24 -plet ($\Sigma$) and a fundamental 5-plet to break the GUT symmetry down to $SU(3)_C\times U(1)_{em}$.
 The Yukawa couplings of the 5-plet Higgs also generate quark and lepton masses. The Higgs sector becomes
 somewhat more complicated in larger GUT structures such as $SO(10)$\cite{georgi} since there is a larger
 symmetry that needs to be broken.   Conventional $SO(10)$ models employ at least two different Higgs
 representations to break the symmetry down to $SU(3)_C\times SU(2)_L\times U(1)_Y$ (a 16 or a 126 to
 change rank, and one of 45, 54 or a 210 to break the symmetry down further \cite{slansky}).
 To achieve electro-weak symmetry breaking and to generate quark and lepton masses
 an additional 10 dimensional representation is also needed.  A minimal $SO(10)$ model  studied recently, for
 example, utilizes one 10, one 126 and one 210 Higgs representations to achieve symmetry breaking and to
 generate masses for the quarks, leptons and the neutrinos\cite{goh}.

 In this paper we discuss the following question: Is it possible to achieve $SO(10)$ symmetry breaking
 all the way down to the $SU(3)_C\times U(1)_{em}$ with a single Higgs representation?
  We find that this is indeed the case if one employs a 144-plet of Higgs of $SO(10)$.
 The 144-plet is contained in the product $10\times 16$, and thus carries one vector and one
 spinor index. An interesting property of the 144-plet which makes such a symmetry breaking
 chain possible is that it contains in it an $SU(5)$ adjoint with a $U(1)$ charge, as well as
 Standard Model Higgs doublet fields. This can be seen from the following decomposition of 144
 under $SU(5)\times U(1)$ subgroup of $SO(10)$
  \beqn\label{144plet}
 144= \bar 5 (3) +5(7) +10(-1) +15(-1) + 24(-5) +40(-1) + \overline{45} (3)
 \label{144}
 \eeqn
 It is significant that the $SU(5)$ adjoint 24(-5) above also carries a U(1) charge. Once the Standard Model
 singlet in it acquires a VEV, it would change the rank of the group, leading to a one-step breaking of
 $SO(10)$ down to $SU(3)_C\times SU(2)_L\times U(1)_Y$. The sub-multiplets
 $\bar 5(3), 5(7)$ and $\overline{45}(3)$  all contain fields with identical quantum
 numbers as the Standard Model Higgs doublet. If one combination of doublets from these sub-multiplets
 is made light by fine tuning, it can be used for electro-weak symmetry breaking. Such fine tuning is
 justified in the context of the multiple vacua of the string landscapes, which has been widely
 discussed recently.
  Although consistency of the first stage of symmetry breaking requires the mass-squared of all the
physical Higgs particles to be positive (including that of the light Higgs doublet), we show that
radiative corrections involving the Higgs self-couplings can turn the mass-squared of the light
Higgs doublet negative, facilitating the second stage of symmetry breaking.

Realistic fermion masses can be obtained  within this minimal
scenario. Recall that the fermions of each family belongs to the 16
dimensional spinor representation of $SO(10)$. Under $SU(5)\times
U(1)$ subgroup the 16 decomposes as follows \beqn\label{16plet}
 16=10(-1) + \bar 5(3) +1(-5) \eeqn
 Fermion masses will arise from quartic couplings  of the $16_i16_j(144 ~144)$ and $16_i16_j$ $(144^*~ 144^*)$.
  These couplings would lead to Dirac masses for all the fermions as well as large Majorana masses for the
 right-handed neutrinos. Since the light Higgs doublet is a linear combination of doublets from 5 and 45 of
 $SU(5)$, the resulting mass pattern is not that of minimal $SU(5)$ and is consistent with experimental data,
 including neutrino oscillations.

The outline of the rest of the paper is as follows: In Sec. 2 we
discuss symmetry and mass growth in the $SU(5)\times U(1)$
language.  In Sec. 3 we discuss the techniques of calculation for
the analysis of $144$ and $\overline{144}$ plet couplings using the method
developed in Ref.\cite{ns}. Here
also we discuss the set of couplings ($144\times \overline{144}$),
$(144\times \overline{144})_1(144\times \overline{144})_1$,
$(144\times \overline{144})_{45}(144\times \overline{144})_{45}$
and $(144\times \overline{144})_{210}(144\times
\overline{144})_{210}$. These couplings are needed in the
computation of symmetry breaking which is then analysed. In Sec. 4
Higgs phenomenon and mass growth are analysed for the breaking of
$SO(10)$. Here it is shown that within the landscape scenario with
fine tuning\cite{ad,lands2} one gets a pair of light Higgs doublets
exactly as in the minimal supersymmetric standard model (MSSM)
while the Higgs triplets and other modes are either absorbed or
become super heavy. In Sec. 5  couplings of quarks and leptons are
discussed and  it is shown that such couplings are quartic in
nature. As an illustration the couplings involving $(16\times
{16})_{10}(144\times {144})_{10}$ and  $(16\times
{16})_{10}(\overline{144}\times \overline{144})_{10}$ are
explicitly discussed.  It is shown that the resulting masses and
mixings are consistent with experimental data.
Conclusions are given in Sec. 6.

 \section{Symmetry breaking and mass  growth with 144 in the ${\bf{SU(5)\times U(1)}}$ language}

 Analysis of the symmetry breaking and of fermion mass
generation with a 144 of Higgs turns out to be rather complicated.
Before we delve into the full detail in the $SO(10)$ language,
which is presented in the next section, we analyze here  these
issues in the simpler $SU(5)\times U(1)$ subgroup language. We
will present our analysis in a non-supersymmetric model.
Generalization to supersymmetry require the addition of a
$\overline{144}$ chiral multiplet, so that the flatness of the
D-term potential can be maintained at the GUT scale, leaving
supersymmetry intact at that scale. The analysis in this section would also hold
for SUSY models with some redefinitions of parameters, provided
that the $144^*$ field is identified with the $\overline{144}$ of
the SUSY $SO(10)$ model.

\subsection{One step GUT symmetry breaking}
Since in the  $SU(5)\times U(1)$ decomposition of $SO(10)$ the 144
contains an $SU(5)$ adjoint carrying a non-zero $U(1)$ charge
(see Eq.(\ref{144plet})), it is instructive to analyze symmetry
breaking of $SU(5)\times U(1)$  with a complex adjoint $\Sigma$.
One can construct such a representation from two  adjoint
representations of $SU(5)$: $\Sigma =\Sigma_1+i\Sigma_2$. Then
$\Sigma$ is not self-adjoint, and we denote the conjugate of
$\Sigma$ as $\Sigma^{\dagger}$.

The most general $SU(5)\times U(1)$ invariant renormalizable
potential involving the $\Sigma$ and $\Sigma^{\dagger}$ fields is
\beqn\label{scalarpotential1}
 V= -M^2 tr(\Sigma \Sigma^{\dagger})
+\frac{\kappa_1}{2} tr(\Sigma^2 \Sigma^{\dagger 2})
 +\frac{\kappa_2}{2} (tr(\Sigma \Sigma^{\dagger}))^2
\nonumber\\
 +\frac{\kappa_3}{2} tr(\Sigma^2) tr(\Sigma^{\dagger 2})
+\frac{\kappa_4}{2} tr(\Sigma \Sigma^{\dagger}  \Sigma \Sigma^{\dagger}   )~.
\eeqn
Among the possible local minima is the one which preserves the Standard Model gauge symmetry  given by the
vacuum structure
 \beqn\label{vev1}
 <\Sigma >= <\Sigma^{\dagger}> =v ~diag(2,2,2, -3, -3)
 \eeqn
 For some range of the parameters of the potential, this minimum will be the global minimum.
 Minimization of the potential gives
 \beqn\label{mincond1}
 v^2 = \frac{ M^2}{ 7(\kappa_1+\kappa_4)+ 30 (\kappa_2 + \kappa_3)}~.
 \eeqn
 Clearly this VEV structure  breaks $SU(5)$ down to $SU(3)_C\times SU(2)_L\times U(1)_Y$.
 Further, since $\Sigma$ is charged under the $U(1)$, its VEV breaks this  $U(1)$. Thus we see that
 the  $SU(5)\times U(1)$ symmetry is broken down to the SM gauge symmetry in one step with one complex
 adjoint Higgs field.  This can also be verified directly by computing the gauge  boson masses.
 The physical Higgs boson masses can all be made positive for some range of parameters of the
 potential. It is then clear that if an $SO(10)$ representation contains sub-multiplets which transform
 under $SU(5)\times U(1)$ symmetry as an adjoint carrying $U(1)$ charge, then there is the possibility
 that this Higgs field can break $SO(10)$ all the way down to the SM gauge symmetry in one step. We observe  that
 the 144-plet of $SO(10)$ is the simplest representation  which has this
 property\footnote{The next simplest possibility is to have a 560 of $SO(10)$, which contains  a 24(-5),
 1(-5), as well as $\bar 5(3), 45(7)$, and $\overline{45}(3)$  under $SU(5)\times U(1)$.}.
Technical details of this assertion in the $SO(10)$ language is
postponed to the next section.

Identical conclusions can be arrived at for the case of
supersymmetric $SU(5)\times U(1)$ model.  The potential of
Eq.(\ref{scalarpotential1}) will become the superpotential (with
$\Sigma^*$ replaced by a chiral superfield $\overline{\Sigma}$), the
couplings $\kappa_i$ will be nonrenormalizable operators with
inverse dimensions of mass, and the mass term $M^2$ will be
replaced by $M$. Thus we conclude that in SUSY $SO(10)$ a
$144+\overline{144}$ pair of chiral superfields can break $SO(10)$
in one step down to the supersymmetric Standard Model gauge group.

\subsection{Electroweak symmetry breaking}
Having recognized that a single 144-plet can break
non-supersymmetric $SO(10)$ down to the SM in one step, we turn
our attention to the subsequent electro-weak symmetry breaking.
As noted earlier, the 144-plet also contains fields which have the
same quantum numbers as the SM Higgs doublet. We explain how these
doublet fragments from the 144-plet can be used for the purpose of
electro-weak symmetry breaking, thus making the model very
economical.

An immediate question that can be raised is how to obtain negative mass-squared for the light Higgs doublet
of the SM arising from the 144-plet. Consistency of the GUT symmetry breaking would require positivity of the
mass-squared of all the physical Higgs bosons, including that of the light SM doublet. If the surviving
symmetry and spectrum below the GUT scale are corresponding to those of the SM, there would be no interactions
that turn the Higgs mass-squared negative needed for electroweak symmetry breaking. In analogy to the stop
squark quartic couplings turning the Higgs mass-squared negative in the supersymmetric SM, we observe that
the quartic couplings between the light Higgs doublet  and any fragment of the 144-plet with mass an order
of magnitude or so below the GUT scale can turn the Higgs doublet mass-squared
negative\footnote{Yukawa couplings between the Higgs doublet and fermions cannot turn the Higgs mass-squared
negative. Cubic self couplings (which are not allowed in the SM Higgs doublet) and/or  quartic/cubic scalar
couplings involving other  fields are necessary for this to happen.}. We illustrate  this mechanism with a
simple $SU(5)$ toy model with an adjoint Higgs field below.

Consider a toy model with global $SU(5)$ symmetry broken
spontaneously by an adjoint Higgs field $\Sigma$. The most general
renormalizable potential for this field is given by
\beqn\label{scalarpotential2}
 V= m^2 Tr(\Sigma^2) +\frac{\kappa_1}{2}
tr(\Sigma^4) + \frac{\kappa_2}{2} (tr(\Sigma^2))^2 + \mu
tr(\Sigma^3). \label{su5} \eeqn One possible VEV structure is as
shown in Eq.(\ref{vev1}). Minimization of the potential of
Eq.(\ref{scalarpotential2}) gives \beqn\label{mincond2}
 m^2=- (7\kappa_1 + 30 \kappa_2)v^2
+\frac{3}{2} \mu v~. \eeqn The $(3,2,-5/6)$ and $(3^*,2,5/6)$
components of $\Sigma$ (under the surviving $SU(3)\times
SU(2)\times U(1)$ symmetry) are Goldstone bosons, while the
$(8,1,0), (1,3,0)$ and the $(1,1,0)$ components have masses given
respectively by \beqn\label{mass}
m_8^2 =\frac{15}{2} \mu v + 5\kappa_1 v^2,\nonumber\\
m_3^2 =-\frac{15}{2} \mu v + 20 \kappa_1 v^2,\nonumber\\
m_1^2 =-\frac{3}{2} \mu v +(14\kappa_1 +60 \kappa_2) v^2.
\eeqn
For a range of parameters, all three squared-masses can be chosen positive, establishing the consistency of
symmetry breaking.

It is possible by fine tuning to make the $SU(2)_L$ triplet field to be much lighter than the $SU(5)$ breaking
scale, while keeping the other two  components heavy. We wish to ask if such a finely tuned triplet can
subsequently break $SU(2)_L$ further down to $U(1)_L$. This would, however, require that  $m_3^2$ turn
 negative at lower  scales even though it starts off as being positive at the high scale. Consider the case when
 $\kappa_1$ and $\mu/v$ are somewhat smaller than one (say of order 0.01), while $\kappa_2$ is of order one.
 Then $m_3$ and $m_8$ are generically an order of magnitude below the GUT scale $v$, while $m_1$ is of order $v$.
 In the momentum range below $v$ and above $m_8$, the  mass parameters $m_3^2$ and $m_8^2$  will evolve, while the
 singlet decouples. The RGE for the running of these mass parameters are found to be
 \begin{eqnarray}\label{rge}
 \frac{dm_8^2}{dt} &=& \frac{1}{8\pi^2} [\frac{15}{8} (\mu +4\kappa_1 v)^2 + \frac{5}{2} (\kappa_1+2\kappa_2)m_8^2
 +\frac{3}{2} \kappa_2 m_3^2]\nonumber\\
  \frac{dm_3^2}{dt} &=& \frac{1}{8\pi^2} [4\kappa_2 m_8^2 + (\frac{1}{2}\kappa_1+\kappa_2) m_3^2]
 \end{eqnarray}
 where $t=log(Q)$. With $\kappa_2$ being of order one and $\kappa_1$, $\mu/v << 1$, we see from these equations that
 if $m_3^2$ at the scale $v$ starts off being smaller than $m_8^2$, it can turn negative in going down from $v$ to the
 mass scale $m_8$. The mass parameter $m_8^2$ will remain positive in this case. This example shows that fine tuning
 of the weak triplet can be done at the scale $m_8$ is such a way that its squared-mass turns  negative at lower
 energy scales.

 In analogy with this example, if any fragment of the 144-plet of $SO(10)$ remains somewhat lighter than the GUT scale,
 then the quartic couplings involving that fragment  and the Higgs doublet would turn the doublet mass-squared negative\footnote{This statement doas not contradicts Michel-Radicati theorem\cite{Michel:1979vv} as we are using radiative correction to turn the doublet mass-squared negative.}.
 It is interesting to note that if such fragments from the 144 that survive below the GUT scale are the color octet(s) and
 the  weak triplet(s), unification of the three SM gauge couplings will occur nicely at  a scale of $(10^{16}-10^{17})$ GeV \cite{Bachas:1995yt}.
 The above analysis and conclusions  are in the context of a non-SUSY $SO(10)$ scenario. In the SUSY $SO(10)$ case the top-stop
 Yukawa couplings will turn the Higgs doublet mass-square negative in the usual way.

\subsection{Doublet-triplet splitting}

We denote the components of the 144 multiplet  by $Q$'s. As can be
seen from Eq.(\ref{144plet}) the Higgs fields reside in the
multiplets $Q_i(\bar 5) + Q^i(5) +Q^i_j(24)
+Q^k_{ij}(\overline{45})$. Similarly we denote the components of
the ${144^*}$ multiplet by $P$'s and in this case the
relevant Higgs multiplets will be $P_i(\bar 5)+ P^i(5) +P^i_j(24)
+P_k^{ij}(45)$. The $\overline{5}, 5$ and $45$  representations contain $SU(2)_L$
doublets and $SU(3)_C$ triplets. Before studying the
doublet-triplet splitting in the full $SO(10)$ theory, here we
analyze this possibility within the simpler $SU(5)\times U(1)$
theory, but with couplings motivated by the full $SO(10)$ theory.
The relevant potential that we consider is given by
\beqn\label{scalarpotential3} V_{DT}= m^2tr(Q_iP^i) + m^2
tr(Q^iP_i) + m^2tr(Q^k_{ij}P_k^{ij})
+\lambda_1 tr(Q_iQ^i) tr(Q^i_jQ^j_i)\nonumber\\
+\lambda_2 tr(P_iP^i) tr(P^i_jP^j_i)
+\lambda_3 tr(Q_iP^i) tr(Q^i_jP^j_i)
+\lambda_3' tr(Q^iP_i) tr(Q^i_jP^j_i)\nonumber\\
 +\lambda_4tr(Q^k_{ij}P^{ij}_k) tr(Q^i_j P^j_i)
+\eta_1 tr(Q_iQ^i Q^j_k Q^k_j)  +\eta_2 tr(P_iP^iP^j_kP^k_j)\nonumber\\
  + \eta_3 tr(Q_iP^iQ^j_kP^k_j)  +\eta_3' tr(Q^iP_iQ^j_kP^k_j) + 
  \eta_4  tr(Q_{ij}^k P^m  Q^i_kP^j_m) + \eta_4' tr (P^{ij}_kP_m P^k_iP^m_j)
  \nonumber\\
  +\eta_5 tr(Q^k_{ij} Q^mQ^i_k Q^j_m) +\eta_5' tr (P^{ij} Q_m P^i_k Q^j_m)  
  +\eta_6 tr(P^{ij}_k Q^k_{lm}Q^l_iP^m_j) 
 \eeqn 
 where $\lambda_i$ and
$\eta_i$ are dimensionless couplings which represent different
types of contractions of indices. It is easily checked that each
term in the potential has an overall zero $U(1)$ quantum number.
There are several  Higgs doublets and Higgs triplets and
anti-triplets in this model. Thus one set of Higgs doublets and
triplets and anti-triplets arise from $Q_i, Q^i,P_i, P^i$.
Specifically the doublets are $Q_{\alpha}, Q^{\alpha}, P_{\alpha},
P^{\alpha}$, while the triplets (anti-triplets) are $Q^a,
P^a$($Q_a, P_a$). Additional Higgs doublets, triplets and
anti-triplets arise from the $45$ of $SU(5)$ (from the
$\overline{144}$-plet) and from the  $\overline{45}$ (from the
144-plet). We discuss the decomposition of these below. The $45$
of $SU(5)$ embedded in $\overline{144}$  has the following
$SU(2)\times SU(3)\times U(1)_Y$ decomposition \beqn\label{45}
45=(2,1)(3) + (1,3)(-2) + (3,3)(-2) + (1,\bar 3) (8)\nonumber\\
 + (2,\bar 3) (-7) +(1,\bar 6)(-2)
+(2,8)(3)
\eeqn One notices that while there is only one $SU(2)$ Higgs
doublet ($\tilde P^{\alpha},\alpha =1,2$), there is one $SU(3)_C$
Higgs triplet $\tilde P^a$ ($a=1,2,3$) and one anti-triplet $\tilde
P_a$. Similarly, for the $\overline{45}$ embedded in $144$ one has
one $SU(2)$ Higgs doublet ($\tilde Q_{\alpha},\alpha =1,2$),  one
$SU(3)_C$ Higgs triplet $\tilde Q^a$ (a=1,2,3) and one
anti-triplet $\tilde Q_a$. The above analysis shows that the Higgs
doublet mass matrix will be $3\times 3$ while the Higgs triplet
mass matrix will be $4\times 4$. We focus first on the Higgs
doublet mass  matrix after $Q^i_j$ and $P^i_j$ develop VEVs. We
display the mass matrix for the  Higgs doublets in the basis where
the rows are ($P_{\alpha}, Q_{\alpha}, h_{\alpha}$) and the
columns by  ($P^{\alpha}, Q^{\alpha}, h^{\alpha}$)

\beqn
 \left[\matrix{
30\lambda_2 v^2 +9\eta_2 v^2  & m^2+30\lambda_3'v^2 +9\eta_3'v^2 & v^2\eta_4' c\cr
 m^2+30\lambda_3v^2 +9\eta_3v^2 & 30\lambda_1 v^2 +9\eta_1 v^2 &  v^2\eta_5' c  \cr
 v^2\eta_4 c   &   v^2\eta_5 c  & m^2+30\lambda_4 v^2+ \frac{21}{4} \eta_6 v^2}\right] \
\label{doubletmass}
\eeqn
where $c=-15\sqrt 3/2\sqrt 2$.
Next, we display the  mass  matrix for the Higgs triplets in the
basis where the rows are labelled ($P_{a}, Q_{a}, \tilde Q_{a},
\tilde P_{a}$) and the columns by  ($P^a, Q^{a}, \tilde P^{a},
\tilde Q^{a} $). In this basis the Higgs triplet mass matrix is
{\small
\beqn
 \left[\matrix{
30\lambda_2 v^2 +4\eta_2 v^2  & m^2+30\lambda_3'v^2 + 4 \eta_3'v^2 & \tilde  c \eta_4' v^2 & 0  \cr
 m^2+30\lambda_3v^2 +4\eta_3v^2 & 30\lambda_1 v^2 +4\eta_1 v^2 &  \tilde  c\eta_5' v^2& 0 \cr
 \tilde c \eta_4 v^2 & \tilde c \eta_5 v^2  & m^2+ 30\lambda_4 v^2 +\eta_6 v^2 &0\cr
0 &  0  &  0 & m^2+30\lambda_4 v^2+ 9\eta_6 v^2}\right] \
\label{tripletmass}
\eeqn
}
where $\tilde c =5\sqrt 2$. 
It is easy to see from the determinants of Eqs.(\ref{doubletmass}) and (\ref{tripletmass}) that one can
arrange for a pair of light Higgs doublets while keeping the Higgs triplets heavy.
 As will be shown in Eqs.(\ref{doubletconstraint}) and
(\ref{tripletconstraint}) from the full $SO(10)$ analysis,we can
identify one of the superposition of the doublet fragments as the
SM Higgs doublet while keeping the Higgs color triplets fragments
all super heavy. It is true that one must fine tune in order to
have the Higgs doublet light, but we find it  very interesting
that with a single 144-plet complete breaking of the $SO(10)$
symmetry down to the residual $SU(3)_C\times U(1)_{em}$ can be
achieved.

\subsection{Fermion mass growth}
For the fermion masses we have the following quartic coupling
allowed by gauge invariance, in terms of $SU(5)\times U(1)$
decomposition \beqn\label{fermionmass}W = \frac{h_1}{M}
10^{ij}10^{kl} \Sigma^j_m Q^m +\frac{h_2}{M} 10^{ij}10^{kl}
\Sigma^n_l Q^m +\frac{h_3}{M}  10^{ij}\bar5_i \Sigma_j^m \bar P_m
+ \frac{h_4}{M} 10^{ij} \bar 5_l \Sigma_i^l\bar P_j
\label{textures} \eeqn From Eq.(\ref{textures}) we see that it is
possible to realize Georgi-Jarlskog type relations with
appropriate choice of the $h_i$ couplings even without the
45 of $SU(5)$ acquiring electroweak VEV.  In general, there are additional
terms involving the 45 of $SU(5)$, which would provide additional
freedom since the Higgs doublet
will now be a linear combination of 5 and 45.

\section{Calculational Techniques for the full SO(10) analysis}

 In this section we discuss the breaking of $SO(10)$ down to
$SU(3)_C\times U(1)_{em}$  in a single step by a single pair of
$144 +\overline{144}$.  Our analysis will be valid for the supersymmetric
$SO(10)$ model as well as for the non--supersymmetric model.  In the
latter case one simply identifies $\overline{144}$ with the $144^*$ field.
However, for formal reasons we consider a
single pair of $160 +\overline{160}$ where the additional $16
+\overline{16}$ that reside in $160 +\overline{160}$ are needed
for consistency as we will see below. The analysis involving the
$144 +\overline{144}$ is rather intricate and we use the
techniques developed recently in Refs\cite{ns} based on the
oscillator method of Refs.\cite{ms,wilczek} to compute the desired
couplings.  There are no cubic couplings of quarks and leptons
with the $144 +\overline{144}$ of Higgs and one needs quartic
couplings to grow quark and lepton masses in this scheme. We
discuss now the basic ingredients of the model. We begin by
displaying  the particle content of $144+\overline{144}$ in
multiplets of $SU(5)$. For $\overline{144}$ plet one has
 \beqn  \label{bar144}
 \overline{144} =\bar 5
({\bf{\cal P}}_i)+5 ({\bf{\cal P}}^i)+ \overline{10}({\bf{\cal
P}}_{ij}) + \overline{15}({\bf{\cal P}}^{(S)}_{ij}) + 24
({\bf{\cal P}}^i_j) + \overline{40}({\bf{\cal P}}^l_{ijk}) +
45({\bf{\cal P}}^{ij}_{k}) \eeqn where the subscripts and
superscripts $i,j,k,..$ are $SU(5)$ indices which take on the
values $1,..,5$. Similarly, for $144$ we find \beqn\label{144}
144= 5 ({\bf{\cal Q}}^i)+\bar 5 ({\bf{\cal Q}}_i)+ 10 ({\bf{\cal
Q}}^{ij}) + 15 ({\bf{\cal Q}}_{(S)}^{ij})+ 24({\bf{\cal Q}}^i_j) +
40({\bf{\cal Q}}^{ijk}_l) + \overline{45}({\bf{\cal Q}}^{i}_{jk})
\eeqn
 To make progress we need a field theoretic description of the
 $144$  and $\overline{144}$ plet of fields.
Possible candidates are the vector-spinors $|\Psi_{(\pm)\mu}>$.
However, an unconstrained vector spinor has $16\times 10=160$ independent
components and is reducible. Thus irreducible vector-spinors with
 dimensionality
144 must be gotten by removing 16 of the 160 components of the
unconstrained vector-spinor. We define the 144-dimensional
constrained vector spinors $|\Upsilon_{(\pm)\mu}>$ by imposition
of the constraint \beq \label{constraint1} \Gamma_{\mu}
|\Upsilon_{(\pm)\mu}> =0 \eeq where $\Gamma_{\mu}$ satisfy a
rank-10 Clifford algebra
\begin{equation}\label{clifford}
\{\Gamma_{\mu},\Gamma_{\nu}\}=2\delta_{\mu\nu}.
\end{equation}
and where $\mu,\nu$ are the $SO(10)$ indices and take on the
values $1,..,10$.

We discuss now further the relationship of $|\Psi_{(\pm)\mu}>$ and $|\Upsilon_{(\pm)\mu}>$.
We begin by writing the 160 and $\overline {160}$ component
spinor:
\begin{equation}\label{160spinor}
|\Psi_{(+)\acute{a}\mu}>=|0>{\bf
P}_{\acute{a}\mu}+\frac{1}{2}b_i^{\dagger}b_j^{\dagger}|0>{\bf
P}_{\acute{a}\mu}^{ij} +\frac{1}{24}\epsilon^{ijklm}b_j^{\dagger}
b_k^{\dagger}b_l^{\dagger}b_m^{\dagger}|0>{\bf P}_{\acute{a}i\mu}
\end{equation}
\begin{equation}\label{bar160spinor}
|\Psi_{(-)\acute{b}\mu}>=b_1^{\dagger}b_2^{\dagger}
b_3^{\dagger}b_4^{\dagger}b_5^{\dagger}|0>{\bf Q}_{\acute{b}\mu}
+\frac{1}{12}\epsilon^{ijklm}b_k^{\dagger}b_l^{\dagger}
b_m^{\dagger}|0>{\bf Q}_{\acute{b}ij\mu}+b_i^{\dagger}|0>{\bf
Q}_{\acute{b}\mu}^i
\end{equation}
where the Latin letters $i,j,k,l,m,...$ are $SU(5)$ indices and
the Greek letters $\mu,\nu,\rho, ...$ represent $SO(10)$ indices.
The Latin subscripts $\acute{a}, \acute{b}, \acute{c},
\acute{d}(=1,2,3)$ are reserved for generation indices. The
reducible fields appearing in Eqs.(\ref{160spinor}) and (\ref{bar160spinor})
can be identified in $SU(5)$ notation as follows \cite{ns}:
\begin{eqnarray}\label{reducible1}
          10=5+{\overline 5}:~~{\bf P}_{\mu}=\left({\bf P}_{c_k},{\bf P}_{\overline c_{k}}\right)\equiv
          \left({\bf P}^{k},{\bf P}_{k}\right)\nonumber\\
{\overline {10}}=5+{\overline 5}:~~{\bf Q}_{\mu}=\left({\bf
Q}_{c_k},{\bf Q}_{\overline c_{k}}\right)\equiv
          \left({\bf Q}^{k},{\bf Q}_{k}\right)
\end{eqnarray}
\begin{eqnarray}\label{reducible2}
100={\overline {50}}+50:~~{\bf P}_{\mu}^{ij}=\left({\bf
P}_{c_k}^{ij},{\bf P}_{\overline c_{k}}^{ij}\right)\equiv
          \left({\bf R}^{[ij]k},{\bf R}_{k}^{[ij]}\right)\nonumber\\
{\overline {100}}={\overline {50}}+50:~~{\bf Q}_{\mu
ij}=\left({\bf Q}_{ijc_k},{\bf Q}_{ij\overline c_{k}}\right)\equiv
          \left({\bf S}^k_{[ij]},{\bf S}_{[ij]k}\right)\nonumber\\
50=45+5:~{\bf R}_k^{[ij]}={\bf
P}_k^{ij}+\frac{1}{4}\left(\delta^j_k\widehat { {\bf
P}}^i-\delta^i_k\widehat {{\bf P}}^j\right)\nonumber\\
{\overline
{50}}={\overline {40}}+{\overline {10}}:~{\bf
R}^{[ij]k}=\epsilon^{ijlmn}{\bf
P}_{lmn}^{k}+\epsilon^{ijklm}\widehat {
{\bf P}}_{lm}\nonumber\\
50=40+10:~{\bf S}_{[ij]k}=\epsilon_{ijlmn}{\bf
Q}^{lmn}_{k}+\epsilon_{ijklm}\widehat { {\bf
Q}}^{lm}\nonumber\\
{\overline {50}}={\overline {45}}+{\overline
{5}}:~{\bf S}^i_{[jk]}={\bf
Q}^i_{jk}+\frac{1}{4}\left({\delta^i_k\widehat {\bf
Q}}_j-{\delta^i_{j}\widehat  {\bf Q}}_k\right)
\end{eqnarray}
\begin{eqnarray}\label{reducible3}
{\overline {50}}=25+{\overline {25}}:~{\bf P}_{i\mu }=\left({\bf
P}_{ic_k},{\bf P}_{i\overline c_{k}}\right)\equiv
          \left({\bf R}^{k}_i,{\bf R}_{ik}\right)~~~~~~~~~~~~~~~\nonumber\\
50=25+25:~{\bf Q}_{\mu }^i=\left({\bf Q}_{\overline c_{k}}^i,{\bf
Q}_{c_k}^i\right)\equiv \left({\bf S}^{i}_k,{\bf
S}^{ik}\right)~~~~~~~~~~~~~~~\nonumber\\
25=24+1:~{\bf R}^i_j={\bf P}^i_j+\frac{1}{5}\delta^i_j{\widehat
{\bf P}},~{\overline {25}}={\overline {10}}+{\overline {15}}:~{\bf
R}_{ij}=\frac{1}{2}\left({\bf P}_{ij}+{\bf
P}_{ij}^{(S)}\right)\nonumber\\
25=24+1:~{\bf S}^i_j={\bf Q}^i_j+\frac{1}{5}\delta^i_j{\widehat
{\bf Q}},~25=10+15:~{\bf S}^{ij}=\frac{1}{2}\left({\bf
Q}^{ij}+{\bf Q}^{ij}_{(S)}\right)
\end{eqnarray}
Further, $b_i^{\dagger}$ and $b_i^{\dagger}$ $(i=1,2,..,5)$ are
the fermionic creation and annihilation operators and obey the
anti-commutation rules\cite{ms}
\begin{equation}\label{commutations}
\{b_i,b_j^{\dagger}\}=\delta_{i}^j;~~~\{b_i,b_j\}=0;
~~~\{b_i^{\dagger},b_j^{\dagger}\}=0
\end{equation}
and the $SU(5)$ singlet state $|0>$ satisfies $b_i|0>=0$. $SO(10)$
invariance requires the following constraints in general
\begin{equation}\label{constraint2}
\Gamma_{\mu}|\Psi_{(+)\mu}>=|\Psi^{\prime}_{(-)}>
\end{equation}
\begin{equation}\label{constraint}
\Gamma_{\mu}|\Psi_{(-)\mu}>=|\Psi^{\prime}_{(+)}>
\end{equation}
where
\begin{eqnarray}\label{psi-'}
|\Psi^{\prime}_{(-)}>=b_1^{\dagger}b_2^{\dagger}
b_3^{\dagger}b_4^{\dagger}b_5^{\dagger}|0>{\widehat {\bf P}}
+\frac{1}{12}\epsilon^{ijklm}b_k^{\dagger}b_l^{\dagger}
b_m^{\dagger}|0>\left({\bf P}_{ij}+6{\widehat {\bf
P}}_{ij}\right)\nonumber\\
+b_i^{\dagger}|0>\left({\bf P}^{i}+{\widehat {\bf P}}^{i}\right)
\end{eqnarray}
\begin{eqnarray}\label{psi+'}
|\Psi^{\prime}_{(+)}>=|0>{\widehat {\bf
P}}+\frac{1}{2}b_i^{\dagger}b_j^{\dagger}|0>\left({\bf
Q}^{ij}+6{\widehat {\bf Q}}^{ij}\right)\nonumber\\
+\frac{1}{24}\epsilon^{ijklm}b_j^{\dagger}
b_k^{\dagger}b_l^{\dagger}b_m^{\dagger}|0>\left({\bf
Q}_{i}+{\widehat {\bf Q}}_{i}\right)
\end{eqnarray}
We note in passing that to get the $144$ and $\overline {144}$
spinors, $|\Upsilon_{(\pm)\mu}>$, we need to impose
Eq.(\ref{constraint1}). This constrain will require setting
$|\Psi^{\prime}_{(\pm)}>=0$ and thus require the following
constraints
\begin{eqnarray}\label{constraint4}
{\widehat {\bf P}}=0,~~~{\widehat {\bf P}}^i=-{\bf
P}^i,~~~{\widehat
{\bf P}}_{ij}=-\frac{1}{6}{\bf P}_{ij}\nonumber\\
{\widehat {\bf Q}}=0,~~~{\widehat {\bf Q}}_i=-{\bf
Q}_i,~~~{\widehat {\bf Q}}^{ij}=-\frac{1}{6}{\bf Q}^{ij}
\end{eqnarray}
Thus we have
 \begin{equation}\label{144from160}
 |\Upsilon_{(\pm)\mu}>=\left(|\Psi_{(\pm)\mu}>\right)_{constraint~of~Eq.(\ref{constraint4})}
\end{equation}
However, as we stated already we will be dealing with the full
$160+\overline{160}$ multiplets.

\noindent To normalize the $SU(5)$ fields contained in the tensor,
$|\Psi_{(\pm)\mu}>$, we carry out a field redefinition
\begin{eqnarray}\label{normalized160}
\{1\}:~~{\widehat {\bf P}}=\sqrt 5{\widehat {\cal
P}},~~~\{{\overline 5}\}:~~{{\bf P}}_i={\cal P}_i,~~~\{5\}:~~{
{\bf P}}^i={\cal P}^i,~~~{\widehat {\bf P}}^i=2{\widehat {\cal
P}}^i\nonumber\\~~~\{\overline {10}\}:~~{ {\bf P}}_{ij}=\sqrt
2{\cal P}_{ij},~~~{\widehat {\bf P}}_{ij}=\frac{1}{2\sqrt
3}{\widehat {\cal P}}_{ij},~~~\{\overline {15}\}:~~{ {\bf
P}}^{(S)}_{ij}=\sqrt{2}{\cal
P}^{(S)}_{ij}\nonumber\\
\{24\}:~~{ {\bf P}}^i_j={\cal P}^i_j,~~~\{\overline {40}\}:~~{
{\bf P}}^{l}_{ijk}=\frac{1}{6}{\cal P}^{l}_{ijk},~~~\{45\}:~~{
{\bf P}}^{ij}_{k}={\cal P}^{ij}_{k}
\end{eqnarray}
\begin{eqnarray}\label{normalizedbar160}
\{1\}:~~{\widehat {\bf Q}}=\sqrt 5{\widehat {\cal Q}},~~~\{{
5}\}:~~{{\bf Q}}^i={\cal Q}^i,~~~\{\overline 5\}:~~{ {\bf
Q}}_i={\cal Q}_i,~~~{\widehat {\bf Q}}_i=2{\widehat {\cal
Q}}_i\nonumber\\~~~\{{10}\}:~~{ {\bf Q}}^{ij}=\sqrt 2{\cal
Q}_{ij},~~~{\widehat {\bf Q}}^{ij}=\frac{1}{2\sqrt 3}{\widehat
{\cal Q}}^{ij},~~~\{{15}\}:~~{ {\bf Q}}_{(S)}^{ij}=\sqrt{2}{\cal
Q}_{(S)}^{ij}\nonumber\\
\{24\}:~~{ {\bf Q}}^i_j={\cal Q}^i_j,~~~\{{40}\}:~~{ {\bf
Q}}_{l}^{ijk}=\frac{1}{6}{\cal
Q}_{l}^{ijk},~~~\{\overline{45}\}:~~{ {\bf Q}}_{ij}^{k}={\cal
Q}_{ij}^{k}
\end{eqnarray}
 In terms of the normalized fields, the kinetic
energy of the $160$ and $\overline {160}$, i.e.,
$-<\partial_A\Psi_{(\pm)\mu}|\partial^A\Psi_{(\pm)\mu}>$, where
$A$ is the Lorentz index,  takes the form
\begin{eqnarray}\label{ke160}
{\mathsf L}^{\overline{160}}_{kin}=-\partial_A{\widehat {\cal
P}}^{\dagger}\partial_A{\widehat {\cal P}}-\partial_A{\cal
P}^{\dagger}_i\partial^A{\cal P}_i-\partial_A{\cal
P}^{i\dagger}\partial^A{\cal P}^i-\partial_A{\widehat{\cal
P}}^{i\dagger}\partial^A{\widehat{\cal
P}}^i\nonumber\\-\frac{1}{2!}\partial_A{\cal
P}_{ij}^{\dagger}\partial^A{\cal P}_{ij}
-\frac{1}{2!}\partial_A{\widehat{\cal
P}}^{\dagger}_{ij}\partial^A{\widehat{\cal
P}}_{ij}-\frac{1}{2!}\partial_A{\cal
P}_{ij}^{(S)\dagger}\partial^A{\cal
P}_{ij}^{(S)}\nonumber\\-\partial_A{\cal
P}^{i\dagger}_j\partial^A{\cal
P}^{i}_j-\frac{1}{3!}\partial_A{\cal
P}^{l\dagger}_{ijk}\partial^A{\cal P}_{ijk}^l
-\frac{1}{2!}\partial_A{\cal P}^{ij\dagger}_{k}\partial^A{\cal
P}^{ij}_k
\end{eqnarray}
\begin{eqnarray}\label{kebar160}
{\mathsf L}^{160}_{kin}=-\partial_A{\widehat {\cal
Q}}^{\dagger}\partial_A{\widehat {\cal Q}}-\partial_A{\cal
Q}^{i\dagger}\partial^A{\cal Q}^i-\partial_A{\cal
Q}^{\dagger}_i\partial^A{\cal Q}_i-\partial_A{\widehat{\cal
Q}}^{\dagger}_i\partial^A{\widehat{\cal
Q}}_i\nonumber\\-\frac{1}{2!}\partial_A{\cal
Q}^{ij\dagger}\partial^A{\cal Q}^{ij}
-\frac{1}{2!}\partial_A{\widehat{\cal
Q}}^{ij\dagger}\partial^A{\widehat{\cal
Q}}^{ij}-\frac{1}{2!}\partial_A{\cal
Q}^{ij\dagger}_{(S)}\partial^A{\cal
Q}^{ij}_{(S)}\nonumber\\-\partial_A{\cal
Q}^{i\dagger}_j\partial^A{\cal
Q}^{i}_j-\frac{1}{3!}\partial_A{\cal
Q}^{ijk\dagger}_{l}\partial^A{\cal Q}_{l}^{ijk}
-\frac{1}{2!}\partial_A{\cal Q}^{k\dagger}_{ij}\partial^A{\cal
Q}^{k}_{ij}
\end{eqnarray}

\section{Symmetry Breaking}
In this section we discuss how $SO(10)$ breaks to
 $SU(3)_C\times SU(2)_L\times U(1)_Y$. For this purpose
we consider a superpotential of the form
\begin{eqnarray}\label{generalsuperpotential}
 {\mathsf W}= M(\overline{160}_H\times 160_H) \nonumber\\+
 \frac{\lambda_1}{M'} (\overline{160}_H\times 160_H)_1
(\overline{160}_H\times 160_H)_1
\nonumber\\
+\frac{\lambda_{45}}{M'} (\overline{160}_H\times 160_H)_{45}
 (\overline{160}_H\times 160_H)_{45}\nonumber\\
+\frac{\lambda_{210}}{M'} (\overline{160}_H\times 160_H)_{210}
 (\overline{160}_H\times 160_H)_{210}
\end{eqnarray}
There are of course many more terms that one can add to
Eq.(\ref{generalsuperpotential}) but we consider only the terms
displayed in Eq.(\ref{generalsuperpotential}) for simplicity.
 The relevant terms in the superpotential that accomplish symmetry breaking
 are
\begin{eqnarray}\label{sbsuperpotential}
   {\mathsf W}_{_{SB}}= M {\bf Q}^i_{\mu}{\bf P}_{i\mu} +\alpha_{_{1}}
{\bf Q}_{\mu}^i{\bf P}_{i\mu}{\bf Q}^j_{\nu}{\bf P}_{j\nu}
+\alpha_{_{2}}{\bf Q}^i_{\mu}{\bf P}_{j\mu}{\bf Q}_{\nu}^j{\bf
P}_{j\nu}
\end{eqnarray}
where
\begin{eqnarray}\label{parameters1}
 \alpha_{_{1}}=\frac{1}{M'} \left(-2\lambda_1-\lambda_{45} +\frac{1}{6}
\lambda_{210}\right)
\nonumber\\
\alpha_{_{2}}=-\frac{1}{M'} \left(4\lambda_{45} +
\lambda_{210}\right)
\end{eqnarray}
 Expanding into the irreducible components we
find
\begin{eqnarray}\label{irredsuperpotential}
W= M {\cal Q}^i_j{\cal P}^j_i + \alpha_{_{1}} {\cal Q}^i_j{\cal
P}^j_i {\cal Q}^k_l{\cal P}^l_k +\alpha_{_{2}}  {\cal Q}^i_k {\cal
P}^k_j {\cal Q}^j_l {\cal P}^l_i + M{\widehat {\cal Q}}{\widehat
{\cal P}} +2\left( \alpha_{_{1}}+\frac{2}{5} \alpha_{_{2}}\right)
 {\cal Q}^k_l {\cal P}^l_k {\widehat {\cal Q}}{\widehat {\cal P}} \nonumber\\
 +\frac{2}{\sqrt 5}\alpha_{_{2}}
{\cal Q}^i_k{\cal P}^k_j {\cal Q}^j_i{\widehat {\cal P}}
+\frac{2}{\sqrt 5} \alpha_{_{1}} {\cal Q}^i_k {\cal P}^k_j {\cal
P}^j_i{\widehat {\cal Q}} +\frac{1}{5} \alpha_{_{2}} {\cal Q}^k_l
{\cal Q}^l_k{\widehat {\cal P}}{\widehat {\cal P}}+\frac{1}{5}
\alpha_{_{1}}{\cal P}^k_l{\cal P}^l_k{\widehat {\cal Q}}{\widehat
{\cal Q}}~~~~~~~~~~~~~
\nonumber\\
+\left(\alpha_{_{1}} +\frac{1}{5} \alpha_{_{2}}\right){\widehat
{\cal Q}}{\widehat {\cal P}}{\widehat {\cal Q}}{\widehat {\cal
P}}~~~~~~~~~~~~~~~~~~~~~~~~~~~~~~~~~~~~~~~~~~~~~~~~~~~~~~~~~~~~~~~~~~~~~
\end{eqnarray}

In the minimization we look for solutions of the type
 \beqn\label{vev2}
  <{\cal Q}^i_j>= q ~diag(2,2,2,-3,-3),~~ <{\cal P}^i_j>= p ~diag(2,2,2,-3,-3)
\eeqn One finds the following results from the minimization of the
potential
\begin{eqnarray}\label{minimization1}
Mp +2p^2q\left(30\alpha_{_{1}} +7 \alpha_{_{2}}\right) +
2\left(\alpha_{_{1}} +\frac{2}{5} \alpha_{_{2}}\right)p
Q_{_{0}}P_{_{0}}
\nonumber\\
+\frac{1}{75}\alpha_{_{2}}q P_{_{0}}^2 + +\frac{26}{5\sqrt
5}\alpha_{_{2}}\left(p Q_{_{0}}+2qP_{_{0}}\right)p =0
\end{eqnarray}
\begin{eqnarray}\label{minimization2}
Mq +2q^2p\left(30\alpha_{_{1}} +7 \alpha_{_{2}}\right) +
2\left(\alpha_{_{1}} +\frac{2}{5} \alpha_{_{2}}\right)q
Q_{_{0}}P_{_{0}}
\nonumber\\
+\frac{1}{75}\alpha_{_{2}}p Q_{_{0}}^2 + +\frac{26}{5\sqrt
5}\alpha_{_{2}}\left(q P_{_{0}}+2pQ_{_{0}}\right)q =0
\end{eqnarray}
\begin{eqnarray}\label{minimization3}
 60 \left[\left(\alpha_{_{1}}+ \frac{2}{5}\alpha_{_{2}}\right) qp+M\right] P_{_{0}}
+\frac{2}{5} \alpha_{_{2}} p^2 Q_{_{0}} +\frac{156}{\sqrt 5}
\alpha_{_{2}}p^2q
\nonumber\\
+2\left( \alpha_{_{1}} +\frac{1}{5}
\alpha_{_{2}}\right)Q_{_{0}}P_{_{0}}^2=0
\end{eqnarray}
and
\begin{eqnarray}\label{minimization4}
 60 \left[\left(\alpha_{_{1}}+ \frac{2}{5}\alpha_{_{2}}\right) qp+M\right] Q_{_{0}}
+\frac{2}{5} \alpha_{_{2}} q^2 P_{_{0}} +\frac{156}{\sqrt 5}
\alpha_{_{2}}pq^2
\nonumber\\
+2\left( \alpha_{_{1}} +\frac{1}{5}
\alpha_{_{2}}\right)Q_{_{0}}^2P_{_{0}}=0
\end{eqnarray}
 where $Q_{_{0}}=<{\widehat {\cal Q}}>$ and
$P_{_{0}}=<{\widehat {\cal P}}>$. The D-flatness condition
$<144>=<\overline{144}>$ gives $q=p$.
With the above vacuum
expectation value (VEV),
 spontaneous breaking occurs so
that $SO(10)\rightarrow SU(3)_C\times SU(2)_L\times U(1)_Y$.
We note that the VEVs $Q_0$ and $P_0$ do not play a role in the
above breakdown as this breakdown will occur even when
$Q_0=0=P_0$.

\section{Higgs Phenomenon and Mass Growth}
We outline here the Higgs phenomenon and the mass growth
associated with the spontaneous breaking given by
Eqs.(\ref{minimization1}-\ref{minimization4}).
 The fields that participate in the Higgs phenomenon include
the 45 vector super multiplet that belongs to the adjoint
representation of $SO(10)$ and the $144+\overline{144}$ chiral
superfields. For the analysis of the Higgs phenomenon it is useful
to decompose the 45-plet of $SO(10)$ in multiplets of $SU(5)$ so
that \beqn \label{45plet} 45=1+10+\overline{10} + 24 \eeqn After
spontaneous symmetry breaking the $10_{45}$ massless vector super
multiplet absorbs the $10_{144}$ chiral multiplet to become a
$10_{45}$ massive vector super multiplet with spins
$(1,\frac{1}{2}, 0)$. Similarly, the $\overline{10}_{45}$ vector
super multiplet absorbs the $\overline{10}_{\overline{144}}$
chiral multiplet to become the $\overline{10}_{45}$ massive vector
super multiplet. Now the 24 plet of $SU(5)$ decomposes under
$SU(3)_C\times SU(2)_L$ as follows \beqn \label{24plet}24=(8,1) +
(\bar 3,2) + (3,2) + (1,3) + (1,1) \eeqn
 After spontaneous breaking the super vector multiplets with the
 quantum numbers $ (\bar 3,2) + (3,2)$ absorb one linear combination
 of the chiral multiplets $((\bar 3,2) + (3,2))_{144}$ and
 $((\bar 3,2) + (3,2))_{\overline{144}}$ becoming  a massive
$ (\bar 3,2) + (3,2)$  vector super multiplets while the orthogonal
linear combination of $((\bar 3,2) + (3,2))_{144}$ and
 $((\bar 3,2) + (3,2))_{\overline{144}}$ which is not absorbed
 becomes massive. The vector super multiplets corresponding
 to $(8,1)+(1,3)+(1,1)$ remain massless. The chiral super multiplets
 corresponding to $(8,1)+(1,3)$ become massive.
 (The  $(1,1)$ components of $24_{144}$ and $24_{\overline{144}}$ require
 special treatment and we return to it below).
 Thus we have accounted
 for the mass growth of the $10+\overline{10}$ vector super multiplet
 and the mass growth of the 12 components $(\bar 3,2) + (3,2)$ of the 24 plet
 vector super multiplet. This leaves us to discuss mass growth of the
 singlet vector super multiplet in Eq.(\ref{45plet}). This mass growth comes
 about by absorption of the chiral superfield combination
$(\frac{2}{5}\Sigma^a_a -\frac{3}{5}\Sigma^{\alpha}_{\alpha})$
where the repeated indices are summed ($a$ is the color index which
takes on values 1,2,3 and $\alpha$ is the $SU(2)$  index and takes
on values 4,5), and $\Sigma^i_j$ is a linear combination of ${\cal
Q}^i_j$ and ${\cal P}^i_j$. Since $\Sigma$ is traceless the above
equals $\Sigma^a_a =-\Sigma^{\alpha}_{\alpha}$. Thus the singlet
vector super multiplet absorbs a linear combination of ${\cal
P}^a_a$ and ${\cal Q}^a_a$ becoming a singlet massive vector
super multiplet while the orthogonal combination of the chiral
superfields ${\cal P}^a_a$ and ${\cal Q}^a_a$ becomes massive.
Thus after the symmetry breaking and Higgs phenomenon only the
$(1,8)+(1,3)+(1,1)$ vector supermultiplet remains massless and the
remaining components of $45$ of the vector super multiplet  become
massive. Similarly, all the unabsorbed components of the
$144+\overline{144}$ become massive. At this stage the gauge group
$SO(10)$ has broken down to
 $SU(3)_C\times SU(2)_L\times U(1)_Y$. To accomplish the
 breaking of the electro-weak symmetry we need a pair of Higgs
 doublets.
 Such a possibility arises for ${\cal Q}_{\alpha}$, ${\cal P}^{\alpha}$.

 To exhibit this we need to compute masses for
  ${\cal Q}_{i}$, ${\cal P}^{i}$. It is also instructive to compute masses
  for ${\cal Q}^{i}$, ${\cal P}_{i}$, ${\widehat{\cal Q}}_{i}$, ${\widehat{\cal
  P}}^{i}$, ${\cal Q}^{i}_{jk}$ and ${\cal P}^{ij}_{k}$
  The relevant terms in the superpotential are
 \begin{eqnarray}\label{doublettripletsuperpotential}
 {\mathsf W}_{_{mass}}= \left[M+\frac{1}{M'}\left(-4\lambda_{1}+6\lambda_{45}-\frac{1}{3}\lambda_{210}\right)<{\bf S}^m_n{\bf R}^n_m> \right]
 \left({\cal Q}_i{\cal P}^i + {\cal Q}^i{\cal P}_i\right) \nonumber\\
-\left[\frac{8}{3}\frac{\lambda_{210}}{M'}<{\bf S}^i_m{\bf
R}^m_j>\right]{\cal Q}_i{\cal P}^j
+\left[\frac{1}{M'}\left(8\lambda_{45}-\frac{2}{3}\lambda_{210}\right)<{\bf
S}^i_m{\bf R}^m_j>\right]
 {\bf S}^l_{[in]}{\bf R}^{[nj]}_l\nonumber\\
  +\left[-\frac{1}{2}M+\frac{1}{M'}\left(2\lambda_{1}+\lambda_{45}-\frac{1}{6}\lambda_{210}\right)<{\bf S}^m_n{\bf R}^n_m> \right]
  {\bf S}_{[ij]}^k{\bf R}_k^{[ij]}
\end{eqnarray}
Eq.(\ref{doublettripletsuperpotential}) makes it apparent why for
technical reasons we need to keep the $160+\overline{160}$
multiplet. To see this let us set all the couplings $\lambda$ to
zero in Eq.(\ref{doublettripletsuperpotential}) so that  the only
terms surviving are proportional to $M$.  Next  suppose we impose
on ${\bf S}_{[ij]}^k$ and ${\bf R}_k^{[ij]}$ the constraint of
Eq.(\ref{constraint4}) so that we are strictly considering only
the $144+\overline{144}$ multiplet. Then we see that the last line
of Eq.(\ref{doublettripletsuperpotential}) contributes an
additional mass term $ -\frac{M}{4}{\cal Q}_i{\cal P}^i$ while
there is no such term for ${\cal Q}^i{\cal P}_i$. This additional
term is clearly not desired and the reason for its appearance is
that we are identifying the $5$ plet in ${\bf R}^{[ij]}_k$ with
${\cal P}^i$ and $\bar 5$ plet in ${\bf S}_{[ij]}^k$ with ${\cal
Q}_i$ because of Eq.(\ref{constraint4}) which feeds in the
undesired additional term.
 Thus for book keeping we must
not impose the constraint of Eq.(\ref{constraint4}) in the
beginning. Returning to Eq.(\ref{doublettripletsuperpotential}),
the corresponding mass terms in the Lagrangian are  given by
\begin{eqnarray}\label{doublettripletlagrangian}
{\mathsf L}= \left|\frac{\partial {\mathsf W}_{_{mass}}}{\partial
{\cal Q}_i}\right|^2+\left|\frac{\partial {\mathsf
W}_{_{mass}}}{\partial {\cal P}^i}\right|^2+\left|\frac{\partial
{\mathsf W}_{_{mass}}}{\partial {\cal
Q}^i}\right|^2+\left|\frac{\partial {\mathsf
W}_{_{mass}}}{\partial {\cal P}_i}\right|^2 +\left|\frac{\partial
{\mathsf W}_{_{mass}}}{\partial {\bf
S}_{[ij]}^k}\right|^2+\left|\frac{\partial {\mathsf
W}_{_{mass}}}{\partial {\bf R}^{[ij]}_k}\right|^2
\end{eqnarray}
Explicitly we have
\begin{eqnarray}\label{explicitlagrangian}
{\mathsf L}=\left|\sigma\right|^2\left({\cal P}_i{\cal
P}_i^{\dagger}+{\cal Q}^i{\cal Q}^{i\dagger}\right)\nonumber\\
+\left|\sigma+\omega_{_{1}}\beta\right|^2\left({\cal
P}^{\alpha}{\cal P}^{\alpha\dagger}+{\cal Q}_{\alpha}{\cal
Q}_{\alpha}^{\dagger}\right)\nonumber\\
+\left|\sigma+\omega_{_{2}}\beta\right|^2\left({\cal P}^{a}{\cal P}^{a\dagger}+{\cal Q}_{a}{\cal Q}_{a}^{\dagger}\right)\nonumber\\
+\frac{1}{2}\left[\left|\rho-\omega_{_{1}}\gamma\right|^2+3
\left|\rho-\frac{1}{2}\left(\omega_{_{1}}+\omega_{_{2}}\right)\gamma\right|^2
\right]\left({\widehat{\cal P}}^{\alpha}{\widehat{\cal
P}}^{\alpha\dagger} +{\widehat{\cal Q}}_{\alpha}{\widehat{\cal
Q}}_{\alpha}^{\dagger}\right)\nonumber\\
+\left[\left|\rho-\omega_{_{1}}\gamma\right|^2+
\left|\rho-\frac{1}{2}\left(\omega_{_{1}}+\omega_{_{2}}\right)\gamma\right|^2
\right]\left({\widehat{\cal P}}^{a}{\widehat{\cal P}}^{a\dagger}
+{\widehat{\cal Q}}_{a}{\widehat{\cal Q}}_{a}^{\dagger}\right)\nonumber\\
+\left|\rho-\omega_{_{1}}\gamma\right|^2\left({\cal
P}_k^{\alpha\beta}{\cal P}_k^{\alpha\beta\dagger}+{\cal
Q}^k_{\alpha\beta}{\cal
Q}^{k\dagger}_{\alpha\beta}\right)\nonumber\\
+\left|\rho-\omega_{_{2}}\gamma\right|^2\left({\cal P}_k^{ab}{\cal
P}_k^{ab\dagger}+{\cal Q}^k_{ab}{\cal
Q}^{k\dagger}_{ab}\right)\nonumber\\
+2\left|\rho-\frac{1}{2}\left(\omega_{_{1}}+\omega_{_{2}}\right)\gamma\right|^2
\left({\cal P}_k^{a\alpha}{\cal P}_k^{a\alpha\dagger}+{\cal
Q}^k_{a\alpha}{\cal Q}^{k\dagger}_{a\alpha}\right)\nonumber\\
+\left[\left|\rho-\frac{1}{2}\left(\omega_{_{1}}+\omega_{_{2}}\right)\gamma\right|^2
-\left|\rho-\omega_{_{1}}\gamma\right|^2 \right]\left({\cal
P}_{\alpha}^{a\alpha}{\widehat{\cal P}}^{a\dagger}+{\cal
Q}^{\alpha}_{a\alpha}{\widehat{\cal
Q}}^{\dagger}_{a}+H.C.\right)\nonumber\\
+\left[\left|\rho-\frac{1}{2}\left(\omega_{_{1}}+\omega_{_{2}}\right)\gamma\right|^2
-\left|\rho-\omega_{_{2}}\gamma\right|^2 \right]\left({\cal
P}_{a}^{\alpha a}{\widehat{\cal P}}^{\alpha\dagger}+{\cal
Q}^{a}_{\alpha a}{\widehat{\cal Q}}^{\dagger}_{\alpha}+H.C.\right)
\end{eqnarray}
where
\begin{eqnarray}\label{parameters2}
\sigma=M+\frac{qp}{M'}\left(-4\lambda_{1}+6\lambda_{45}-\frac{1}{3}\lambda_{210}\right)\left(30+\frac{P_{_{0}}Q_{_{0}}}{pq}\right)\nonumber\\
\rho=-\frac{1}{2}M+\frac{qp}{M'}\left(2\lambda_{1}+\lambda_{45}-\frac{1}{6}\lambda_{210}\right)\left(30+\frac{P_{_{0}}Q_{_{0}}}{pq}\right)\nonumber\\
\omega_{_{1}}=qp\left[9-\frac{3}{\sqrt 5}\left(\frac{P_{_{0}}}{p}+\frac{Q_{_{0}}}{q}\right)+\frac{1}{5}\frac{P_{_{0}}Q_{_{0}}}{pq}\right]\nonumber\\
\omega_{_{2}}=qp\left[4+\frac{2}{\sqrt 5}\left(\frac{P_{_{0}}}{p}+\frac{Q_{_{0}}}{q}\right)+\frac{1}{5}\frac{P_{_{0}}Q_{_{0}}}{pq}\right]\nonumber\\
\gamma=\frac{1}{M'}\left(8\lambda
_{45}-\frac{2}{3}\lambda_{210}\right)\nonumber\\
\beta=-\frac{8}{3}\frac{\lambda_{210}}{M'}
\end{eqnarray}
 The masses for the fields ${\cal P}^i, {\cal Q}_i$,
${\cal Q}^i, {\cal P}_i$,
  ${\cal Q}^k_{ij}$, and ${\cal P}^{ij}_k$ can be computed from Eqs.(\ref{minimization1}-\ref{minimization4}) and
  (\ref{explicitlagrangian}). It is
easily checked that the masses vanish unless one has a
non-vanishing $\lambda_{45}$ or $\lambda_{210}$. A scrutiny of the
mass growth above shows that the masses of the Higgs doublets
${\cal Q}_{\alpha},{\cal P}^{\alpha}$ are split from the Higgs
triplets ${\cal Q}_{a}, {\cal P}^{a}$. Indeed this allows one to
fine tune the masses of the Higgs doublets to zero by the
condition
\begin{eqnarray}
\frac{MM'}{qp}+  \left(-120\lambda_1 + 180
\lambda_{45}\right)\left(1+\frac{1}{30}\frac{P_{_{0}}Q_{_{0}}}{pq}\right)\nonumber\\
+
\lambda_{210}\left[-34+\frac{8}{\sqrt
5}\left(\frac{P_{_{0}}}{p}+\frac{Q_{_{0}}}{q}\right)-\frac{13}{15}\frac{P_{_{0}}Q_{_{0}}}{pq}\right]=0
\label{doubletconstraint}
\end{eqnarray}
 With this constraint one finds that the Higgs doublets ${\cal
Q}_{\alpha}, {\cal P}^{\alpha}$ are massless while the Higgs
triplets ${\cal Q}_{a}, {\cal P}^{a}$ are massive. Thus the Higgs
triplet mass $M_{H_3}$ of the fields ${\cal Q}_a, {\cal P}^a$,
under the constraint that the Higgs doublet $ {\cal Q}_{\alpha},
{\cal P}^{\alpha}$ be massless, is given by \beq M_{H_3} =
\frac{40}{3}\frac{qp}{M'}\left[1-\frac{1}{\sqrt
5}\left(\frac{P_{_{0}}}{p}+\frac{Q_{_{0}}}{q}\right)\right]\lambda_{210}
\label{tripletconstraint}
\eeq We note that it is not possible to achieve a doublet-triplet
splitting for the multiplets ${\cal Q}^i$ and ${\cal P}_i$.  Thus
the doublet-triplet splitting we are considering is unique.
Further, we find that ${\cal Q}^i$, ${\cal P}_i$ develop a mass
\begin{eqnarray}\label{massQsuperiPsubi}
M_{{\cal Q}^i,{\cal P}_i}=24\frac{qp}{M'}\left[1-\frac{1}{3\sqrt
5}\left(\frac{P_{_{0}}}{p}+\frac{Q_{_{0}}}{q}\right)+\frac{1}{45}\frac{P_{_{0}}Q_{_{0}}}{pq}\right]\lambda_{210}
\end{eqnarray}
In the above $Q_0$ and $P_0$ are crucial in getting a pair of light
Higgs doublets. Thus suppose we  have $Q_0=0=P_0$
 in Eqs.(\ref{minimization1}-\ref{minimization4}). In this case one  finds an additional constraint.
 Thus from Eqs.(\ref{minimization3}) and (\ref{minimization4}) one finds that $Q_0=0=P_0$ imply that
 $\alpha_2=0$. Under this constraint Eq.(\ref{doubletconstraint}) is not consistent
 with Eqs.(\ref{minimization1}) and (\ref{minimization2}). i.e., one cannot find a pair of light
 Higgs  doublets. We note that this result is a consequence of considering
 only a limited number of couplings in Eq.(\ref{generalsuperpotential}). Inclusion of a larger set
 of couplings should allow one to get  consistent solutions without inclusion
 of the singlet VEVs.

 To
summarize the results thus far we have here a complete breaking of
$SO(10)$ to $SU(3)_C\times SU(2)_L\times U(1)_Y$. To get a pair of
light Higgs we need to invoke a string landscape scenario which
has been extensively discussed recently\cite{ad,lands2}. Effectively in
this framework we use a fine tuning to keep one pair of Higgs
doublets light while keeping the Higgs triplets heavy.
 With the usual radiative
electroweak symmetry breaking mechanism, the light doublets of
Higgs can develop VEVs breaking the $SU(2)_L\times U(1)_Y$
symmetry down to $U(1)_{em}$. Thus with the above mechanism one
can break $SO(10)$ down to $SU(3)_C\times U(1)_{em}$ with just one
pair of $144 + \overline{144}$ of Higgs.
 A similar analysis
shows that one gets masses for the remaining parts of the $144
+\overline{144}$ chiral  multiplets not absorbed by the vector
bosons which become heavy. The heavy spectrum of this model
differs significantly from the standard $SU(5)$ and the standard
$SO(10)$ models. It consists of several parts: (i) the super heavy
lepto-quarks associated with the breaking of the
$SO(10)\rightarrow SU(3)_C\times SU(2)_L\times U(1)_Y$, (ii) the
super heavy Higgs triplet field, (iii) the components of 24 plet
fields ${\cal P}^i_j, {\cal Q}^i_j$ which are unabsorbed by the
Higgs phenomenon and become super heavy, (iv) the remaining
components of $144+\overline{144}$, aside from a 24 plets of
$SU(5)$ each in $144$ and $\overline{144}$ discussed in (iii) and
excluding the Higgs doublets ${\cal Q}_{\alpha}, {\cal
P}^{\alpha}$, which become super heavy, and (v) $16+\overline{16}$
plet of super heavy fields. Gauge coupling unification will
require a careful analysis of contribution of each of the above.
We do not address this question further here.

We discuss briefly the issue of split vs non-split supersymmetry
breaking. Eq.(\ref{doubletconstraint}) is a tree level relation
and its imposition
 to achive light Higgs doublets presumes that the loop corrections
to the Higgs masses are small. Specifically it requires that the
scale of supersymmetry breaking is not high, e.g.,  the masses of
of squarks are at the electroweak scale and not at the GUT scale,
 An alternative possibility is that one
may impose Eq.(\ref{doubletconstraint}) but with loop correction
included. In this case we can allow for split supersymmetry
scenario where the masses of the squarks are superheavy
while gaugino masses may lie at the electroweak scale. Thus the
model we are considering can accommodate
 a split or a non-split supersymmetry breaking scenario.

\section{Couplings of Quarks and Leptons with 144 and ${\bf\overline{144}}$
of Higgs }

The $144$ and $\overline{144}$ plets of Higgs have no $SO(10)$
invariant trilinear coupling with the 16 plet of matter. However
one can write quartic couplings of the type $({1}/{M_P})(16\times
16)(144\times 144)$ and $({1}/{M_P})(16\times
16)(\overline{144}\times \overline{144})$, where $M_P$ is a super
heavy mass. These interactions will generate effective cubic
couplings with the light Higgs doublets ${\cal Q}_{\alpha}, {\cal
P}^{\alpha}$ after spontaneous breaking of $SO(10)$ discussed in
Secs.3 and 4.  The size of these couplings is $O(M_G/M_P)$. The
most general set of quartic couplings involving quarks and leptons
are\footnote{We have not included the 
$(16\times 16)_{120}({144}\times{144})_{120}$ and 
$(16\times 16)_{{120}}
(\overline{144}\times \overline{144})_{120}$
couplings here since these couplings are anti-symmetric in the
generation indices and vanish with only a single $144 +\overline{144}$.} 

 \beqn \label{quarkleptonsuperpotential} (16\times
16)_{10}(144\times 144)_{10}, ~~~ (16\times
16)_{10}(\overline{144}
\times \overline{144})_{10},\nonumber\\
(16\times 16)_{\overline{126}} (144\times 144)_{126},~~~
(16\times 16)_{\overline{126}}
(\overline{144}\times \overline{144})_{126} \eeqn
The couplings
${\left(16\times16\right)_{10}\left(144\times144\right)_{10}}$ and
${\left(16\times16\right)_{10}
\left(\overline{144}\times\overline{144}\right)_{10}}$  arise from
the following structures
\begin{eqnarray}\label{10superpotential}
{\mathsf W}^{(10)}=\frac{1}{2}\Phi_{\nu\cal U}{\cal
M}^{^{(10)}}_{{\cal U}{\cal U}'}\Phi_{\nu{\cal
U}'}+h^{^{(10)}}_{\acute{a}\acute{b}}<\Upsilon^{*}_{(+)\acute{a}\mu}|B\Gamma_{\nu}
|\Upsilon_{(+)\acute{b}\mu}>k_{_{{\cal U}}}^{^{(10)}}\Phi_{\nu\cal
U}\nonumber\\
+f^{^{(10)}}_{\acute{a}\acute{b}}<\Psi^{*}_{(+)\acute{a}}|B\Gamma_{\nu}
|\Psi_{(+)\acute{b}}>l_{_{{\cal U}}}^{^{(10)}}\Phi_{\nu\cal U} +
+\bar{h}^{^{(10)}}_{\acute{a}\acute{b}}<\Upsilon^{*}_{(-)\acute{a}\mu}|B\Gamma_{\nu}
|\Upsilon_{(-)\acute{b}\mu}>\bar{k}_{_{{\cal
U}}}^{^{(10)}}\Phi_{\nu\cal U}
\end{eqnarray}
where the indices ${\cal U}, {\cal U}'$ run over several Higgs
representations of the same kind, ${\cal M}^{^{(10)}}$ represents
the mass matrix and $f^{^{(10)}}$, $k^{^{(10)}}$,
$\bar{k}^{^{(10)}}$, and $l^{^{(10)}}$ are constants. $B$ is the
usual $SO(10)$ charge conjugation operator defined by
$B=\prod_{\mu =odd} \Gamma_{\mu}$. The semi-spinors $\Psi_{(\pm)}$
transforms  as a 16($\overline{16}$)-dimensional irreducible
representation of $SO(10)$ and contain
$1+\overline{5}+10$($1+5+\overline{10}$) in its $SU(5)$
decomposition. $|\Psi_{(+)\acute{a}}>$ is given by
\begin{equation}\label{16dimspinor}
|\Psi_{(+)\acute{a}}>=|0>{\bf
M}_{\acute{a}}+\frac{1}{2}b_i^{\dagger}b_j^{\dagger}|0>{\bf
M}_{\acute{a}}^{ij} +\frac{1}{24}\epsilon^{ijklm}b_j^{\dagger}
b_k^{\dagger}b_l^{\dagger}b_m^{\dagger}|0>{\bf M}_{\acute{a}i}
\end{equation}
Elimination of the super heavy fields $\Phi_{\nu\cal U}$
 using the F-flatness condition gives
\begin{eqnarray}\label{dim5}
{\mathsf W}_{dim-5}^{(10)}= {\mathsf
W}^{^{^{{(16\times16)}_{10}{(144\times144)}_{10}}}} +{\mathsf
W}^{^{^{{(16\times16)}_{10}{(\overline{144}\times\overline{144})}_{10}}}}
\end{eqnarray}
where
\begin{eqnarray}\label{dim5bar144}
{\mathsf W}^{^{^{{(16\times16)}_{10}{(\overline{144}\times
\overline{144})}_{10}}}}=
-2\xi_{\acute{a}\acute{b},\acute{c}\acute{d}}^{^{(10)}}
<\Psi^{*}_{(+)\acute{a}}|B\Gamma_{\rho}|\Psi_{(+)\acute{b}}><\Upsilon^{*}_{(+)\acute{c}\nu}|\Gamma_{\rho}
|\Upsilon_{(+)\acute{d}\nu}>\nonumber\\
=-4\xi_{\acute{a}\acute{b},\acute{c}\acute{d}}^{^{(10)}}
[<\Psi^{*}_{(+)\acute{a}}|Bb_i|\Psi_{(+)\acute{b}}><\Upsilon^{*}_{(+)\acute{c}\nu}|Bb_i^{\dagger}
|\Upsilon_{(+)\acute{d}\nu}>\nonumber\\
+<\Psi^{*}_{(+)\acute{a}}|Bb_i^{\dagger}|\Psi_{(+)\acute{b}}><\Upsilon^{*}_{(+)\acute{c}\nu}|Bb_i
|\Upsilon_{(+)\acute{d}\nu}>]\nonumber\\
=2\xi_{\acute{a}\acute{b},\acute{c}\acute{d}}^{^{(10)(+)}}
[\epsilon_{jklmn}{\bf M}_{\acute{a}i}^{\bf T}{\bf
M}_{\acute{b}}^{ij}{\bf P}_{\acute{c}\mu}^{kl\bf T}{\bf
P}_{\acute{d}\mu}^{mn}-8{\bf M}_{\acute{a}i}^{\bf T}{\bf
M}_{\acute{b}}^{ij}{\bf P}_{\acute{c}j\mu}^{\bf T}{\bf
P}_{\acute{d}\mu}\nonumber\\
+\epsilon_{jklmn}{\bf M}_{\acute{a}}^{kl\bf T}{\bf
M}_{\acute{b}}^{mn}{\bf P}_{\acute{c}i\mu}^{\bf T}{\bf
P}_{\acute{d}\mu}^{ij}-8{\bf M}_{\acute{a}j}^{\bf T}{\bf
M}_{\acute{b}}{\bf P}_{\acute{c}i\mu}^{\bf T}{\bf
P}_{\acute{d}\mu}^{ij}]
\end{eqnarray}
and
\begin{eqnarray}\label{dim5144}
{\mathsf W}^{^{^{{(16\times16)}_{10}{(144\times 144)}_{10}}}}=
-2\zeta_{\acute{a}\acute{b},\acute{c}\acute{d}}^{^{(10)}}
<\Psi^{*}_{(+)\acute{a}}|B\Gamma_{\rho}|\Psi_{(+)\acute{b}}><\Upsilon^{*}_{(-)\acute{c}\nu}|B\Gamma_{\rho}
|\Upsilon_{(-)\acute{d}\nu}>\nonumber\\
=-4\zeta_{\acute{a}\acute{b},\acute{c}\acute{d}}^{^{(10)}}
[<\Psi^{*}_{(+)\acute{a}}|Bb_i|\Psi_{(+)\acute{b}}><\Upsilon^{*}_{(-)\acute{c}\nu}|Bb_i^{\dagger}
|\Upsilon_{(-)\acute{d}\nu}>\nonumber\\
+<\Psi^{*}_{(+)\acute{a}}|Bb_i^{\dagger}|\Psi_{(+)\acute{b}}><\Upsilon^{*}_{(-)\acute{c}\nu}|Bb_i
|\Upsilon_{(-)\acute{d}\nu}>]\nonumber\\
=2\zeta_{\acute{a}\acute{b},\acute{c}\acute{d}}^{^{(10)(+)}}
[8{\bf M}_{\acute{a}i}^{\bf T}{\bf M}_{\acute{b}}^{ij}{\bf
Q}_{\acute{c}\mu}^{k\bf T}{\bf Q}_{\acute{d}kj\mu} -8{\bf
M}_{\acute{a}i}^{\bf T}{\bf M}_{\acute{b}}{\bf
Q}_{\acute{c}\mu}^{i\bf T}{\bf Q}_{\acute{d}\mu}\nonumber\\
 -{\bf M}_{\acute{a}}^{ij\bf T}{\bf M}_{\acute{b}}^{kl}{\bf
Q}_{\acute{c}kl\mu}^{\bf T}{\bf Q}_{\acute{d}ij\mu} +{\bf
M}_{\acute{a}}^{ij\bf T}{\bf M}_{\acute{b}}^{kl}{\bf
Q}_{\acute{c}ik\mu}^{\bf T}{\bf Q}_{\acute{d}jl\mu}\nonumber\\
-{\bf M}_{\acute{a}}^{ij\bf T}{\bf M}_{\acute{b}}^{kl}{\bf
Q}_{\acute{c}il\mu}^{\bf T}{\bf Q}_{\acute{d}jk\mu}+
\epsilon^{ijklm}{\bf M}_{\acute{a}i}^{\bf T}{\bf
M}_{\acute{b}}{\bf Q}_{\acute{c}jk\mu}^{\bf T}{\bf
Q}_{\acute{d}lm\mu}\nonumber\\
+\epsilon_{ijklm}{\bf M}_{\acute{a}}^{ij\bf T}{\bf
M}_{\acute{b}}^{kl}{\bf Q}_{\acute{c}\mu}^{m\bf T}{\bf
Q}_{\acute{d}\mu}]
\end{eqnarray}
and where
\begin{eqnarray}\label{parameters3}
\xi_{\acute{a}\acute{b},\acute{c}\acute{d}}^{^{(10)}}=
f_{\acute{a}\acute{b}}^{^{(10)}}h_{\acute{c}\acute{d}}^{^{(10)}}l_{_{{\cal
U}}}^{^{(10)}}\widetilde{{\cal M}}^{^{(10)}}_{{\cal U}{\cal
U}'}k_{_{{\cal U}'}}^{^{(10)}}
\nonumber\\
\zeta_{\acute{a}\acute{b},\acute{c}\acute{d}}^{^{(10)}}=
f_{\acute{a}\acute{b}}^{^{(10)}}\bar{h}_{\acute{c}\acute{d}}^{^{(10)}}l_{_{{\cal
U}}}^{^{(10)}}\widetilde{{\cal M}}^{^{(10)}}_{{\cal U}{\cal U}'}\bar{k}_{_{{\cal U}'}}^{^{(10)}}
\nonumber\\
\widetilde{{\cal M}}^{^{(10)}}=\left[{\cal
M}^{^{(10)}}+\left({\cal M}^{^{(10)}}\right)^{\bf {T}}\right]^{-1}
\end{eqnarray}
We note that $\xi_{\acute{a}\acute{b}
,\acute{c}\acute{d}}^{^{(10)(+)}}$, and
$\zeta_{\acute{a}\acute{b},\acute{c}\acute{d}}^{^{(10)(+)}}$, are
the same as defined by Eq.(\ref{parameters3}), provided we replace
$h$'s, $\bar{h}$'s, $f$'s by $h^{^{(+)}}$'s, $\bar{h}^{^{(+)}}$'s,
$f^{^{(+)}}$'s respectively where (+) indicates that the couplings
are symmetric, i.e.,
$h^{^{{(10)(\pm)}}}_{\acute{a}\acute{b}}=\frac{1}{2}\left(h^{^{(10)}}_{\acute{a}\acute{b}}\pm
h^{^{(10)}}_{\acute{b}\acute{a}}\right)$. \noindent The above
couplings produce quark and lepton masses after GUT symmetry
breaking followed by spontaneous breaking of the electroweak
symmetry. A preliminary analysis shows that the relation on
Yukawa couplings such as $h_b=h_t$ does not hold. It
would be interesting to study the quark-lepton
textures\cite{g,n,bb} in this framework. Further, the preliminary
analysis shows that baryon and lepton number violating dimension
five operators in this theory are rather different than what one
has in the usual $SU(5)$ and $SO(10)$
models\cite{pdecay1,pdecay2,pdecay3,pdecay4,suppress}. Thus, for
example, in $SU(5)$ unified models the baryon and lepton number
violating dimension five operators arise from the Higgs triplet
exchange from pairs of $5+\bar 5$. In the present model there are
several sources of baryon and lepton number violations, including
Higgs triplets from $5$ and $\bar 5$ and
 from the exchange of $45+\overline{45}$ present in
$144+\overline{144}$ of Higgs.
A full analysis of this issue involves
additional couplings where the mediation occurs via
 120, 126 etc and is beyond the scope of this paper.
 An analysis of this will be given elsewhere\cite{bgns}.
We note here that if in Eq. (53) only couplings involving $\bar{144}$ are kept,
the resulting fermion mass matrices will have the same structure as in \cite{goh}
with a single 10 and one $\overline{126}$ coupling to fermions.
Such matrices are fully consistent with experimental data.

\section{Conclusion}
In conclusion we have investigated a new class of
$SO(10)$ models where the breaking of $SO(10)$ down to
$SU(3)_C\times SU(2)_L\times U(1)_Y$ can occur  in a single
step. Further, it is possible  to achieve with fine tuning,
 a pair of light Higgs doublets which is justifiable within a string based
landscaped scenario.
 The light Higgs doublets allow one to break
the electroweak symmetry $SU(2)_L\times U(1)_Y$ down to
$U(1)_{em}$ and thus $SO(10)$ can break to $SU(3)_C\times
U(1)_{em}$. The cubic interactions of the light Higgs doublets
with quarks and leptons arise from quartic interactions of the
type $(16.16)(144.144)$ and
$(16.16)(\overline{144}.\overline{144})$ after spontaneous
breaking of $SO(10)$. In this scenario
  the baryon and lepton number violating dimension five operators receive
  contributions not just
from the conventional Higgs triplet fields but also from the
exchange of $45+\overline{45}$ components of $144+\overline{144}$.
The above feature distinguishes the above scenario from the
conventional models and would lead to different estimates
on the proton lifetime and in conventional GUTs. A more
detailed analysis of the quark -lepton masses as well as of
the proton life time is outside the scope of this paper
and will be dealt with elsewhere\cite{bgns}.
Finally we note that above the GUT scale one has a large number of degrees of
freedom and the renormalization group evolution is very rapid and thus
the theory becomes nonperturbative. We view this theory as descending
directly from a theory of quantum gravity where the scale of
quantum gravity (e.g., string scale) may lie close to the GUT scale.

\begin{center}
{\bf ACKNOWLEDGEMENTS}
\end{center}
The work of KB is supported in part by DOE grant DE-FG03-98ER-41076.
The work of IG is supported in part by NSF grant PHY-0098791.
The work of PN and RS is supported in part by NSF grant  PHY-0139967.
Part of this work was done while PN was visting the Max Planck Institute, Munich.
He acknowledges support from the Alexander von Humboldt Foundation and
thanks the MPI for the hospitality extended him.


\begin{thebibliography}{999}

\bibitem{georgi}
H. Georgi, in Particles and Fields (edited by C.E. Carlson), A.I.P.,
1975; H. Fritzch and P. Minkowski, Ann. Phys. {\bf 93}, 193 (1975).

\bibitem{slansky}
R.~Slansky,
Phys.\ Rept.\  {\bf 79}, 1 (1981).


\bibitem{goh}
  K.~S.~Babu and R.~N.~Mohapatra,
  Phys.\ Rev.\ Lett.\  {\bf 70}, 2845 (1993);
  B.~Bajc, G.~Senjanovic and F.~Vissani,
  Phys.\ Rev.\ Lett.\  {\bf 90}, 051802 (2003);
H.~S.~Goh, R.~N.~Mohapatra and S.~Nasri,
Phys.\ Rev.\ D {\bf 70}, 075022 (2004);
  S.~Bertolini, M.~Frigerio and M.~Malinsky,
  Phys.\ Rev.\ D {\bf 70}, 095002 (2004);
  K.~S.~Babu and C.~Macesanu,
  arXiv:hep-ph/0505200.



\bibitem{ns}
P.~Nath and R.~M.~Syed,
Phys.\ Lett.\ B {\bf 506}, 68 (2001);
Nucl.\ Phys.\ B {\bf 618}, 138 (2001);
Nucl.\ Phys.\ B {\bf 676}, 64 (2004).

\bibitem{Bachas:1995yt}
  C.~Bachas, C.~Fabre and T.~Yanagida,
  Phys.\ Lett.\ B {\bf 370}, 49 (1996);
  J.~L.~Chkareuli and I.~G.~Gogoladze,
  Phys.\ Rev.\ D {\bf 58}, 055011 (1998).

\bibitem{Michel:1979vv}
  L.~Michel,
CERN-TH-2716
{\it Contribution to Colloq. on Fundamental Interactions, in honor of Antoine Visconti,
 Marseille, France, Jul 5-6,} 1979;
 L.~Michel and L.~Radicati, Ann. Phys. (NY) {\bf 66}, 758 (1971).


\bibitem{ms}
R.N. Mohapatra and B. Sakita, Phys. Rev. {\bf D21}, 1062 (1980).

\bibitem{wilczek}
F. Wilczek and A. Zee, Phys. Rev. {\bf D25}, 553 (1982).

\bibitem{ad}
S.~Kachru, R.~Kallosh, A.~Linde and S.~P.~Trivedi,
Phys.\ Rev.\ D {\bf 68}, 046005 (2003);
F.~Denef and M.~R.~Douglas,
JHEP {\bf 0405}, 072 (2004);
N.~Arkani-Hamed and S.~Dimopoulos,
arXiv:hep-th/0405159; M.~Dine, E.~Gorbatov and S.~Thomas,
arXiv:hep-th/0407043; I.~Antoniadis and S.~Dimopoulos,
hep-th/0411032.



\bibitem{lands2}
 B.~Kors and P.~Nath,
  Nucl.\ Phys.\ B {\bf 711}, 112 (2005);
 K.~S.~Babu, T.~Enkhbat and B.~Mukhopadhyaya,
  arXiv:hep-ph/0501079;
 K.~R.~Dienes, E.~Dudas and T.~Gherghetta,
  arXiv:hep-th/0412185.




\bibitem{g}
H.~Georgi and C.~Jarlskog,
Phys.\ Lett.\ B {\bf 86}, 297 (1979).

\bibitem{n}
P.~Nath,
Phys.\ Rev.\ Lett.\  {\bf 76}, 2218 (1996);
Phys.\ Lett.\ B {\bf 381}, 147 (1996).

\bibitem{bb}
K.~S.~Babu and S.~M.~Barr,
Phys.\ Lett.\ B {\bf 381}, 137 (1996).

\bibitem{pdecay1}
S. Weinberg, Phys. Rev. {\bf D26}, 287 (1982);
N. Sakai and T. Yanagida, Nucl. Phys. {\bf B197}, 533 (1982);
S. Dimopoulos, S. Raby and F. Wilczek, Phys. Lett. {\bf B112}, 133
(1982); J. Ellis, D.V. Nanopoulos and S. Rudaz, Nucl. Phys. {\bf B202},
 43 (1982).

\bibitem{pdecay2}
P.~Nath, A.~H.~Chamseddine and R.~Arnowitt,
Phys.\ Rev.\ D {\bf 32}, 2348 (1985);
 J. Hisano, H. Murayama and T. Yanagida, Nucl. Phys. {\bf B402}, 46 (1993);
T. Goto, T. Nihei and J. Arafune, Phys. Rev. {\bf D52}, 505 (1995);
K.~S.~Babu and S.~M.~Barr,
Phys.\ Lett.\ B {\bf 381}, 137 (1996);
P. Nath and R. Arnowitt,
Phys.\ Atom.\ Nucl.\  {\bf 61}, 975 (1998);
T. Goto and T. Nihei, Phys. Rev. {\bf D59}, 115009 (1999);
K.~S.~Babu and M.~J.~Strassler,
arXiv:hep-ph/9808447;
B.~Bajc, P.~Fileviez Perez and G.~Senjanovic,
Phys.\ Rev.\ D {\bf 66}, 075005 (2002).

\bibitem{pdecay3}
V. Lucas and S. Raby,  Phys. Rev. {\bf D54}, 2261 (1996); Phys. Rev. {\bf D55}, 6986
(1997).

\bibitem{pdecay4}
  K.S. Babu, J.C. Pati and F. Wilczek, Nucl. Phys. {\bf B566} 33 (2000);
Phys.\ Lett.\ B {\bf 423}, 337 (1998).

\bibitem{suppress}
K.~S.~Babu and S.~M.~Barr,
Phys.\ Rev.\ D {\bf 48}, 5354 (1993);
Z. Chacko and R.N. Mohapatra, Phys.Rev. {\bf D59}, 011702 (1999);
Phys. Rev. Lett. {\bf 82}, 2836 (1999);
Z. Berezhiani, Z. Tavartkiladze and M. Vysotsky,
hep-ph/9809301;
I.~Gogoladze and A.~Kobakhidze,
Phys.\ Atom.\ Nucl.\  {\bf 60} (1997) 126
[Yad.\ Fiz.\  {\bf 60N1} (1997) 136];
G. Altarelli, F. Feruglio, and I. Masina,
JHEP {\bf 0011}, 040 (2000);
T.~Ibrahim and P.~Nath,
Phys.\ Rev.\ D {\bf 62}, 095001 (2000);
Q. Shafi and Z. Tavartkiladze, Phys. Lett. {\bf B487}, 145 (2000);
N. Maekawa, Prog. Theor. Phys. {\bf 106}, 401 (2001);
K.~Turzynski,
JHEP {\bf 0210} (2002) 044.

\bibitem{bgns}
K.~S. ~Babu, I.~ Gogoladze, P. ~Nath, and R. ~M. ~ Syed,
in progress.

\end{thebibliography}
\end{document}